\documentclass[prc,twocolumn,showpacs,preprintnumbers,amsmath,amssymb,nofootinbib]{revtex4}

\usepackage{graphicx}
\usepackage{dcolumn}
\usepackage{bm}
\usepackage[mathlines]{lineno}
\usepackage{multirow}
\usepackage{CJK}
\usepackage{epstopdf}

\newcommand{\rr} {\boldsymbol{r}}

\begin{document}
\begin{CJK*}{GBK}{song}
\title{ Global analysis of Skyrme forces with the higher-order density dependencies}

\author{Zhi-Wei Zuo }
\author{Jun-Chen Pei }
\email{peij@pku.edu.cn}

\author{Xue-Yu Xiong }

\author{Yi Zhu }

\affiliation{State Key Laboratory of Nuclear
Physics and Technology, School of Physics, Peking University,  Beijing 100871, China}

\begin{abstract}

The density dependent term in Skyrme forces is essential, which simulates three-body and many-body correlations
beyond the low-momentum two-body interaction.
We speculate that a single density term may be insufficient and
a higher-order density dependent term is added. The present work
investigates the influences of higher-order density dependencies based on extended
UNEDF0 and SkM$^{*}$ forces. The global descriptions of nuclear masses and charge radii have been presented.
  Consequently the extended UNEDF0 force
 gives a global rms error on binding energies of 1.29 MeV.
 The influences on fission barriers and equation of state have also been investigated.
The perspectives to improve
Skyrme forces have also been discussed, including global center-of-mass corrections
and Lipkin-Nogami pairing corrections.

\end{abstract}


\maketitle
\end{CJK*}

\section{Introduction}\label{secInt}
The Skyrme force~\cite{skyrme} is a widely used non-relativistic phenomenological low-momentum effective nuclear force.
The success of the Skyrme force is mainly attributed to its inclusion of a density dependent
term, which becomes a state-dependent in-medium interaction and simulates
 three-body and many-body correlations in the self-consistent mean-field framework.
It was known that other bare two-body density-independent forces can not describe simultaneously nuclear binding energies
and charge radii~\cite{negele}.
The standard Skyrme forces adopt a single density dependent term.
Practical calculations involve a wide range of densities~\cite{benderrev}, from dilute densities at nuclear surface halos to very high densities in neutron stars,
 and thus a single density
dependent term may be insufficient. A natural way to extend the
Skyrme force is to add an additional higher-order density dependent term.
This is consistent with the order-by-order expansion of the energy density functional of atomic gases~\cite{lee-yang}.
The pionless effective field theory results in a similar expression with the Skyrme energy density functional~\cite{furnstahl}, which provides another clue for a higher-order density-dependent term.

In a previous study~\cite{xiong}, we have investigated the influences of the higher-order density dependency based on
the SLy4 force~\cite{sly4}. We demonstrated that the extended SLy4 force can generally improve
the descriptions of binding energies, by reducing the rms error of global binding energies from 2.9 MeV to 2.3 MeV. The high-order density
dependency can also impact the equation of state at very high densities.
It is desirable to further investigate the influences based on other Skyrme forces,
and to study the general behaviors of the  higher-order density dependency.

The fast calculations of the entire nuclear landscape now days enable us to explore the
optimizations of effective nuclear forces from different perspectives.
In recent years,  it is worth to mention that there are many developments to improve nuclear energy density functionals
such as UNEDF~\cite{markus1,markus2,markus3}, Fyans-DFT~\cite{fyans}, BCPM~\cite{bcpm}, SeaLL1-DFT~\cite{bulgac}, QMC-DFT~\cite{stone16}, Gogny-HFB~\cite{d1m}, Brussels-DFT~\cite{hfb22}, and  covariant-DFT~\cite{pwzhao}.
It is still a challenge to develop a highly-accurate universal nuclear energy density functional for bulk properties and dynamics.
Thus different Skyrme parameterizations have been developed to emphasize the description accuracies of nuclear masses~\cite{markus1}, fission barriers~\cite{markus2} and shell structures~\cite{markus3}, respectively.
The Bayesian analysis and covariant analysis to study correlations between parameters,
and correlations between parameters and physical observables can provide useful information for optimizations~\cite{jdm}.
Despite the statistical analysis, the detailed studies of local fluctuations in the global description is also desirable
to identify physics at specific nuclear mass regions.

On the other hand, there are also many efforts to go beyond the standard Skyrme force or beyond the Hartree-Fock approximation.
There are quite some initiatives to construct nuclear energy density functional from effective field theory and ab initio perspectives~\cite{furnstahl,grasso,jacek,duguet,mario}.
In addition, it is known that the unrestricted Hartree-Fock framework naturally breaks all symmetries to taken into account many-body correlations to some extent~\cite{ring,benderrev,uhf}.
The broken symmetries can be restored via projection techniques, which would bring more correlations and beyond-mean-field corrections~\cite{uhf}.
Indeed the inclusion of collective correlation energies can significantly improve the descriptions of nuclear masses~\cite{zpli,d1m,klupfel}.
In this respect, the optimization of Skyrme forces including various restoration corrections
should be systematically explored.

In this work, we have investigated the
global descriptions of nuclear masses and charge radii and the influences of an additional higher-order density dependent term.
There have been various extensions of Skyrme-type energy density functionals~\cite{17,18,19,20,cochet,agrawal06}, however, their advantages are not clear and applications are very limited.
We intend to perform fine optimizations to evaluate the prospects of the extended Skyrme forces.
Our studies are based on two very different Skyrme forces: UNEDF0~\cite{markus1} and SkM$^{*}$~\cite{skm}.
UNEDF0 is best optimized for nuclear masses with a rms error of 1.455 MeV ~\cite{markus1} and SkM$^{*}$ is very successful for fission barriers~\cite{abaran}.
Then we optimize the extended UNEDF0 and SkM$^{*}$ forces and investigate their performances in various aspects.
Furthermore, we studied the global center-of-mass corrections and Lipkin-Nogami pairing corrections.
These corrections are approximate restorations corresponding to the translational symmetry and the non-conservation of particle numbers, respectively.
The detailed global analysis of these corrections is useful for the development of high-precision nuclear energy density functionals, which
is our ultimate goal.

\begin{table*}[bhpt]
\begin{center}

  \caption{ The refitted parameters of the extended Skyrme forces based on UNEDF0 and SkM$^{*}$ forces. The units for $t_0$, $t_3$ and $t_{3E}$
  are $\rm{MeV\cdot fm^3}$,  $\rm{MeV\cdot fm^{3(1+\gamma)}}$ and $\rm{MeV\cdot fm^{3(\gamma+\frac{4}{3})}}$, respectively. Other parameters have not been adjusted.   }
  \label{table1}
  \centering

  \begin{tabular}{lcccccc}
  \hline
    &UNEDF0           & UNEDF0$_{\rm ext1}$ &  UNEDF0$_{\rm ext2}$   &  SkM$^*$        &SkM$^{*}_{\rm ext1}$       &SkM$^{*}_{\rm ext2}$   \\
\hline
$ t_{0}$             & -1883.6878  &-2007.948  & -2140.306  &-2645.0     &-2035.587  & -2325.478  \\
$ t_{3}$             & 13901.948    &11616.664  & 13869.309  &15595.0      &8007.383  &11608.668 \\
$ t_{3E}$            & 0  &3216.9303    &  1402.674         &0      &4795.359  &2534.788   \\
$ x_0 $               & 0.00974  &-0.0494  &  -0.2363    &0.09        &0.2376    &0.2358  \\
$ x_3 $                & -0.3808 &-0.4722 & -0.7760     &0       &-0.07488   &0.2720   \\
$ x_{3E} $             &  0    &-0.1540  & 1.5051    &0        &0.9955  &-0.4692    \\
$\gamma$    &0.3219      & $\frac{1}{4}$ & $\frac{1}{4}$    & $\frac{1}{6}$    & $\frac{1}{6} $  & $\frac{1}{6}$                     \\
\hline

\end{tabular}
\end{center}
\end{table*}

\section{Theoretical framework }\label{secformula}

Systematic calculations in this work are based on the self-consistent deformed Skyrme-Hartree-Fock+BCS method.
The Hartree-Fock equation is solved by the SKYAX code in axial-symmetric coordinate-spaces~\cite{skyax}.
Considering the possible shape coexistences in some nuclei, calculations with different initial deformations have been
performed. The Skyrme force includes the standard two-body interactions $v_{ij}^{(2)}$ and the density-dependent two-body interactions $v_{ij}^{(2)'}$ as,
\begin{eqnarray}
V_{\rm Skyrme}={\sum_{i<j}v_{ij}^{(2)}}+{\sum_{i<j}v_{ij}^{(2)'}}
\end{eqnarray}
The standard two-body density-independent term can be written as~\cite{ring},
\begin{equation}
\begin{array}{ll}
{v_{ij}^{(2)}}=&\displaystyle t_{0}(1+x_{0}P_{\sigma})\delta({\rr}_i-{\rr}_j)\vspace{5pt}\\
&+\displaystyle \frac{1}{2}t_{1}(1+x_1P_{\sigma})[\delta({\rr}_i-{\rr}_j){\bf{k}}^2+{\bf{k'}}^2\delta({\rr}_i-{\rr}_j)] \vspace{5pt}\\
&+\displaystyle t_2(1+x_2P_{\sigma}){\bf{k'}}\cdot\delta({\rr}_i-{\rr}_j){\bf{k}} \vspace{5pt}\\
&+\displaystyle iW_0(\sigma_i+\sigma_j)\cdot{\bf{k'}}\times\delta({\rr}_i-{\rr}_j){\bf{k}}
\end{array}
\label{skyrme}
\end{equation}
In contrast to the standard spin-orbit term in SkM$^{*}$~\cite{skm}, the spin-orbit term in UNEDF0 has been extended by including an explicit isovector degree of freedom~\cite{markus1}.
The extended density dependent two-body interaction includes two terms, with
a density dependency power factor  $\gamma$ and a higher-order power factor $\gamma + \frac{1}{3}$,
\begin{equation}
\begin{array}{ll}
\mathop{v_{ij}^{(2)'}} = & \displaystyle \frac{1}{6}t_3(1+x_3P_{\sigma})\rho(\textbf{R})^{\gamma}\delta({\rr}_i-{\rr}_j)  \vspace{5pt} \\
& + \displaystyle \frac{1}{6}t_{3E}(1+x_{3E}P_{\sigma})\rho(\textbf{R})^{\gamma+\frac{1}{3}}\delta({\rr}_i-{\rr}_j)  \\
\end{array}
\label{extend}
\end{equation}

In the above equations, $t_i$, $x_i$ are standard Skyrme parameters.
In the extended forces, we introduced 2 more additional parameters $t_{3E}$ and $x_{3E}$.
In the SLy4 force~\cite{sly4} and SkM$^{*}$~\cite{skm}, the power factor $\gamma$ takes $1/6$ and then we consider the next higher order power of $1/2=1/6+1/3$.
In the UNEDF forces, the power factors  $\gamma$ are around 1/3. It is useful to explore different combinations of density dependencies.

In the BCS calculations, the mixed pairing interaction is adopted as~\cite{mixed}:
\begin{equation}
V_{\rm mix}(\rr, \rr')=v_{\rm p,n}(1-\frac{\rho(\rr)}{2\rho_0})\delta({\rr}-{\rr'}) ,
\end{equation}
where $v_{p,n}$ is the pairing strengthes for protons and neutrons, respectively, and $\rho_0$ is
the saturation density 0.16 $fm^{-3}$. In the BCS scheme, a smooth pairing cutoff has been adopted~\cite{bender-p}.
 The pairing strengthes are given in Table ~\ref{table2}.

Next we refit the extended Skyrme parameters for finite nuclei with the Simulated Annealing Method~\cite{sam}.
The fitting procedure has been described in our previous work~\cite{xiong}.
We only refit the momentum-independent parameters, $t_0$, $t_3$, $t_{3E}$, $x_0$, $x_3$, $x_{3E}$,
and keep other parameters unchanged. The momentum independent parameters are correlated via the $s$-wave channel. Indeed,  $t_0$, $t_3$ are directly related through regularization as
the leading-order terms for nuclear saturation properties~\cite{yang}.
In this way the influences of the extended higher-order density dependent term can be clearly illustrated.
The optimization would be more reliable as fewer parameters are adjusted. It is expected that the
future optimization of all parameters including momentum dependent terms
can further improve the descriptions.
The fitting procedure takes into account
binding energies of 50 nuclei across the landscape and  charge radii of 8 spherical nuclei.

\begin{figure}[!htbp]
  \begin{center}
  \includegraphics[width=0.48\textwidth]{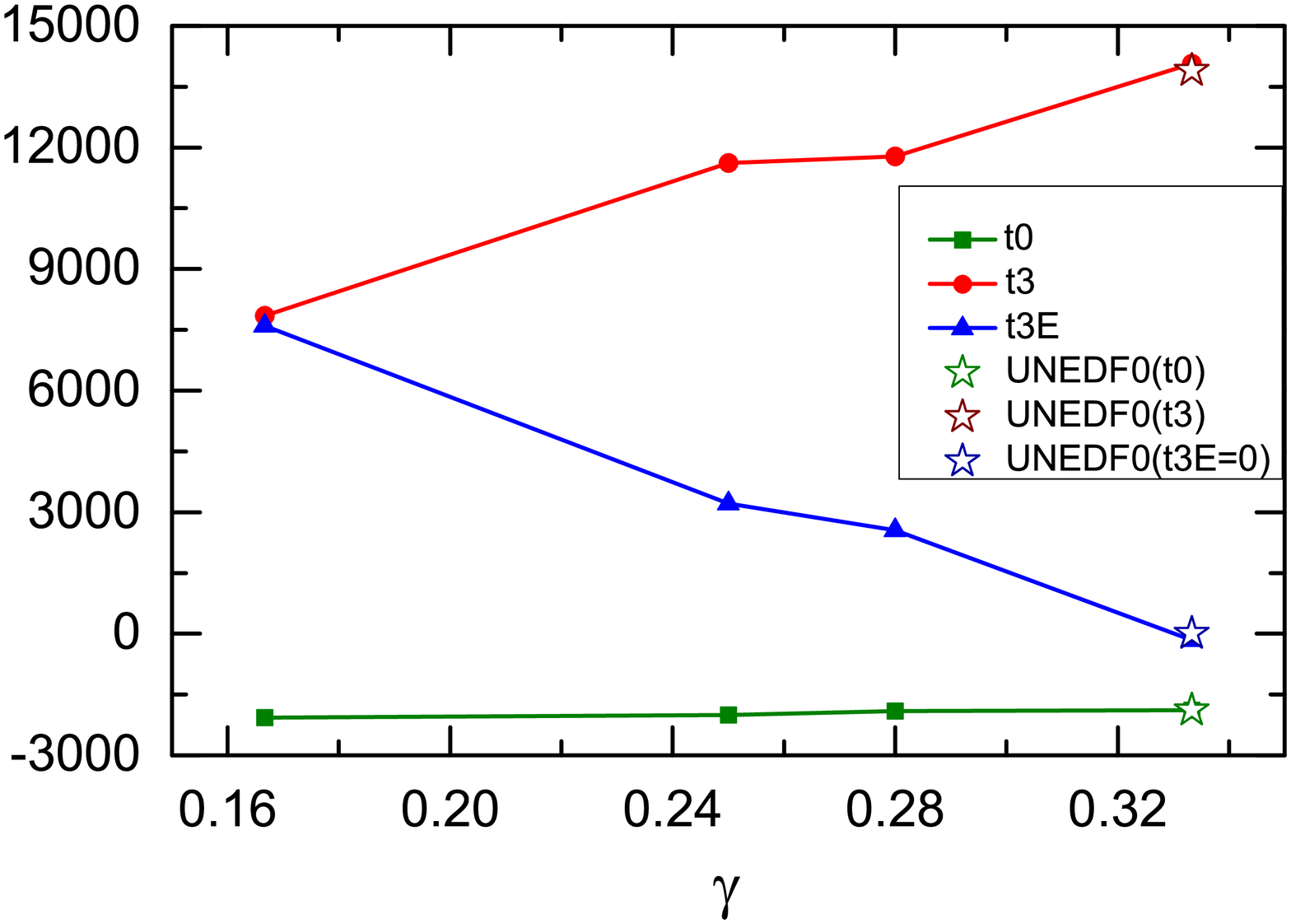}\\
  \caption{(Color online) The refitted parameters: $t_0$, $t_3$ and $t_{3E}$ of the extended UNEDF0 force are given as function of the power factor $\gamma$.
  The parameters of the original UNEDF0 force are also given as stars.  }
  \label{fig-t3}
  \end{center}
\end{figure}

\section{Results and discussions}\label{secRsl}
\subsection{The extended parametrizations}
There have been various density dependencies in Skyrme forces~\cite{stone}. The power factor $\gamma$ in the
Skyrme force is also a parameter, ranging from $1/6$ to 1. We studied the relation between the parameters
$t_0$, $t_3$ and $t_{3E}$ based on the UNEDF0 force by varying $\gamma$.
 In Fig.\ref{fig-t3}, with increasing
$\gamma$, $t_3$ increases and $t_{3E}$ decreases. We see
that $t_0$ is slightly increased with increasing $\gamma$. If $\gamma$ adopts $1/3$, the obtained $t_{3E}$ becomes slightly negative as -160.389.
In the original UNEDF0, $\gamma$ is 0.3219, and in this case we say that $t_{3E}$ is zero, being consistent with
the systematic behavior of $t_{3E}$.  This trend indicates that even higher-order density dependent terms would become negative,
as also obtained in Ref.~\cite{agrawal}. In Fig.\ref{fig-t3}, with a very small $\gamma$, on the other hand, the $t_{3E}$ term would be dominated and the $t_3$ term would be reduced.
  Note that this trend is obtained by fitting finite nuclei.
UNEDF0 is obtained by the optimization of  the free parameter $\gamma$ and then the $t_{3E}$ term is eliminated.
In the extended UNEDF0 force, we have another parameter $x_{3E}$ and this additional isospin degree of freedom can improve the
Skyrme force, as demonstrated in the following. In contrast to SLy4 and SkM$^{*}$ forces having $\gamma$=1/6, UNEDF0 with $\gamma$ around 1/3 has
   little room for a higher-order density dependent term.  By comparing the values of $t_3$ and $t_{3E}$, we assume that in a reasonable combination,  $t_{3E}$, as a higher-order term,  should be smaller than $t_{3}$. Therefore we adopt
$\gamma$ as $1/4$ to refit the extendend UNEDF0 force, in which the power factor of the higher-order term is $1/4+1/3$=$7/12$.  The optimized
extended Skyrme forces are given in Table~\ref{table1}.

\begin{figure}[t]
\begin{center}
  \includegraphics[width=0.48\textwidth]{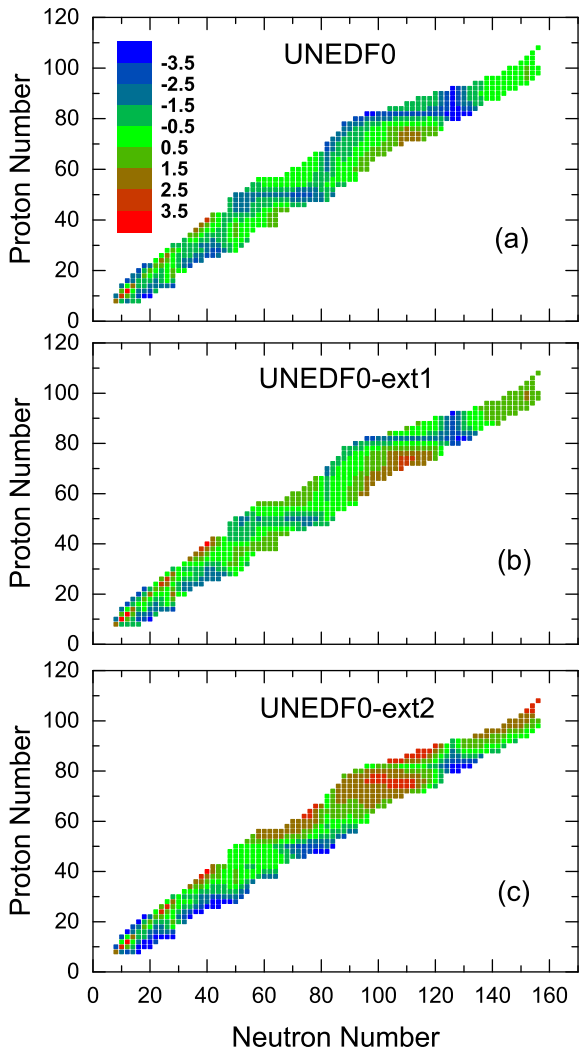}\\
  \caption{(Color online) The binding energy differences between theoretical calculations and experimental data~\cite{audi}
  for 603 even-even nuclei, given as $E_B^{\rm Calc.}-E_B^{\rm Expt.}$ in MeV. The results are obtained by Hartree-Fock+BCS calculations with (a) UNEDF0, (b) UNEDF0$_{\rm ext1}$, (c) UNEDF0$_{\rm ext2}$. See Table \ref{table1} for the parameter sets.}
  \label{fig-unedf}
\end{center}
\end{figure}

\begin{figure}[t]
  \begin{center}
  \includegraphics[width=0.48\textwidth]{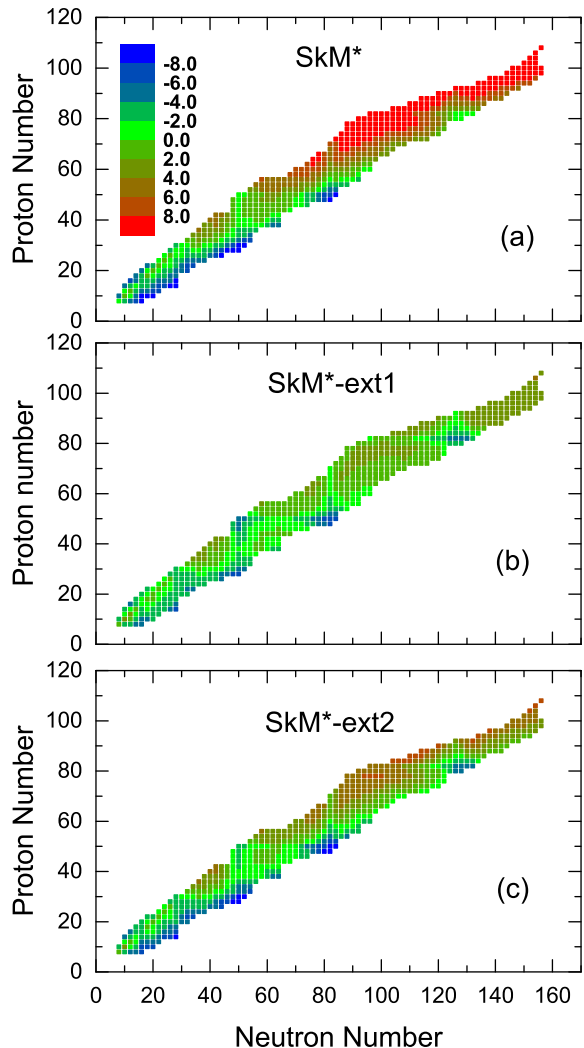}\\
  \caption{(Color online)  The binding energy differences between calculations and experiments~\cite{audi}
  for 603 even-even nuclei, given as $E_B^{\rm Calc.}-E_B^{\rm Expt.}$ in MeV. The results are obtained by Hartree-Fock+BCS calculations with (a) SkM$^{*}$, (b)  SkM$^{*}_{\rm ext1}$, (c)  SkM$^{*}_{\rm ext2}$. See Table \ref{table1} for the parameter sets. }
  \label{fig-skm}
\end{center}
\end{figure}

\begin{figure}[t]
  \begin{center}
  \includegraphics[width=0.48\textwidth]{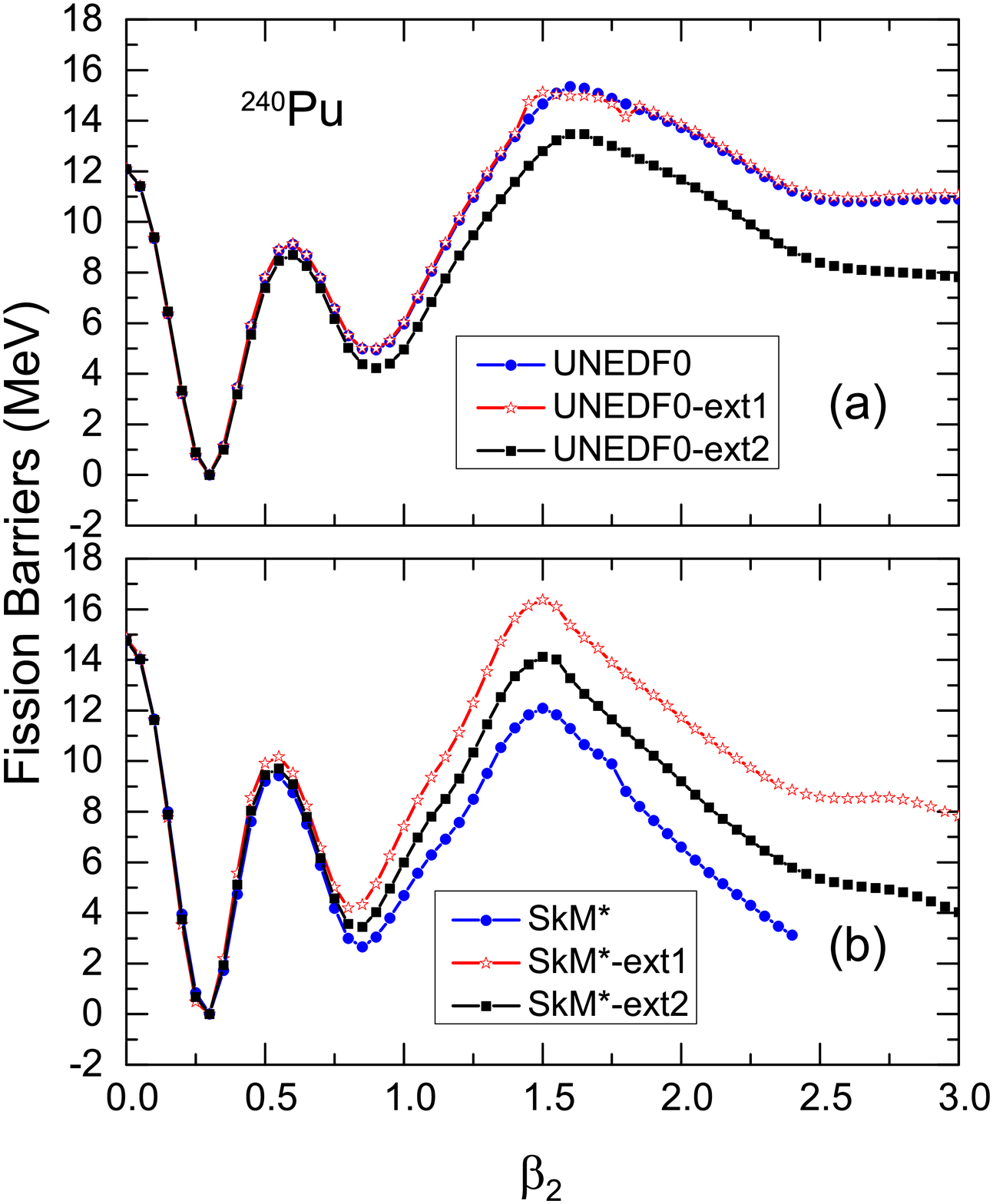}\\
  \caption{(Color online) Calculated symmetric fission barriers of $^{240}$Pu as a function of quadrupole deformation $\beta_2$ with extended Skyrme forces.
  (a) with UNEDF0, UNEDF0$_{\rm ext1}$, UNEDF0$_{\rm ext2}$ forces. (b) with SkM$^{*}$, SkM$^{*}_{\rm ext1}$, SkM$^{*}_{\rm ext2}$ forces.
  The parameters are listed in Table \ref{table1}.
  }
  \label{fig-fission}
  \end{center}
\end{figure}

\subsection{Global binding energies}

To evaluate the influences of the additional density dependent term, we refit
the extended UNEDF0 force. With the optimized UNEDF0$_{\rm ext1}$ force, we did global studies
of nuclear ground state properties based on Skyrme-Hartree-Fock+BCS calculations.  Fig.\ref{fig-unedf} displays the binding energy differences between theoretical calculations
and experimental data
of 603 even-even known nuclei. UNEDF0 has been optimized in a large scale for binding energies, with
a global rms error of 1.455 MeV for 520 even-even nuclei~\cite{markus1}. This is the best description of binding energies with the standard Skyrme force.
In our Hartree-Fock+BCS calculations of 603 nuclei, the global rms values of UNEDF0 and UNEDF0$_{\rm ext1}$ are 1.503 and 1.316 MeV, respectively.
We see the additional higher-order density dependent term can reduce the rms by 12$\%$. For the region $A\leqslant$ 80, the rms is 1.58 MeV.
For the region $A>$ 80, the rms is 1.23 MeV. We do not refit the UNEDF0 force since it has been extremely optimized~\cite{markus1}.
In the region of heavy nuclei, we see the discrepancies between theoretical and experimental values
are dominated by the overestimated shell effects. The UNEDF0$_{\rm ext1}$ has slightly adjusted
the balance between the $^{208}$Pb region and the deformed neutron-rich region around $^{178}$Yb, $^{182}$Hf and $^{186}$W.  In light nuclei,
one of the main discrepancies are from the $N=Z$ nuclei. We see that both theoretical calculations
 generally underestimate the binding energies of these self-conjugated light nuclei and overestimate the binding energies of drip-line light nuclei.
 This can be explained as
 clustering effects and $np$-correlations are absent in the Skyrme-Hartree-Fock framework.  Indeed the valance $n$-$p$ interactions of $N=Z$ light nuclei have
 been remarkably underestimated by the nuclear density functional theory~\cite{stoitsov07}.

Figure \ref{fig-skm} displays the global studies of binding energies of SkM$^{*}$ and extended SkM$^{*}$ forces.
The SkM$^{*}$ force~\cite{skm} has been widely used for fission studies due to its small surface-energy coefficient.
In Fig.\ref{fig-skm}(a), we see the SkM$^{*}$ force is not good at descriptions of global binding energies,
and the rms of binding energies is 6.305 MeV.  It overestimates the binding energies of neutron-rich light and medium nuclei
and underestimates the binding energies of proton-rich heavy and superheavy nuclei.

In Fig.\ref{fig-skm}(b), we refit SkM$^{*}$ with the the higher-order density dependent term as SkM$^{*}_{\rm ext1}$.
We see that the SkM$^{*}_{\rm ext1}$ descriptions of binding energies have been much improved with a rms error of 2.358 MeV.
We see again similar features between SkM$^{*}_{\rm ext1}$ and UNEDF0$_{\rm ext1}$ results.
The binding energies of light neutron-rich nuclei are overestimated and the binding energies of some $N=Z$ nuclei are underestimated.
Then the predicted neutron drip-line in the light and medium mass region could be overextended by the SkM$^{*}$ force.

\subsection{Fission barriers}

It has been a long-standing goal to simultaneously and accurately describe nuclear masses and fission barriers.
It is known that SkM$^{*}$ is good at descriptions of fission barriers and UNEDF0 is good at descriptions of nuclear masses.
In Fig.~\ref{fig-unedf}(c) and Fig.\ref{fig-skm}(c), we refit the  extended UNEDF0 and SkM$^{*}$ with the input of the fission isomer energy of $^{240}$Pu.
The isomer energy is defined by the binding energy difference between the ground state and the fission isomer.
The obtained extended Skyrme forces are UNEDF0$_{\rm ext2}$ and SkM$^{*}_{\rm ext2}$, respectively, as listed in Table~\ref{table1}.
We see that the UNEDF0$_{\rm ext2}$ and SkM$^{*}_{\rm ext2}$ descriptions of binding energies become worse again with rms errors of 1.87 MeV and 3.676 MeV, respectively.
This demonstrates that there is a competition in the simultaneous optimization of nuclear masses and fission barriers (or surface properties) .
We see that SkM$^{*}$ , SkM$^{*}_{\rm ext2}$ and UNEDF0$_{\rm ext2}$ forces which have been fitted with fission barriers,  have all significantly underestimated binding energies of proton-rich heavy nuclei and overestimated binding energies of neutron-rich light nuclei,
implying conflicting isospin-dependent corrections on binding energies and fission barriers.
Such a competition has also been reflected in the increased rms values of 1.91 MeV of UNEDF1~\cite{markus2} compared to the 1.455 MeV of UNEDF0~\cite{markus1}, in which the optimization of UNEDF1 includes
both fission isomers and nuclear masses while the optimization of UNEDF0 doesn't include
fission isomers.

Figure \ref{fig-fission} displays the calculated symmetric fission barriers of $^{240}$Pu with the extended UNEDF0 and SkM$^{*}$ forces.
The triaxial deformation and reflection-asymmetric deformation have
not been considered, which are important for descriptions of fission barriers.
In principle, multi-dimensional constraint calculations should be performed to study fission barriers~\cite{lu2012,lu2014}.
Nevertheless it is suitable to consider the fission isomer energies in the symmetric case to constrain large deformation properties to reduce computing time.
The experimental excitation energy of the fission isomer of $^{240}$Pu is 2.8 MeV~\cite{pu240}.
In our cases, the fission isomer energies of $^{240}$Pu calculated by UNEDF0, UNEDF0$_{\rm ext1}$ and UNEDF0$_{\rm ext2}$ are 4.95, 4.99, 4.23 MeV, respectively.
The fission isomer energies of $^{240}$Pu calculated by SkM$^{*}$, SkM$^{*}_{\rm ext1}$ and SkM$^{*}_{\rm ext2}$ are 2.65, 4.2, 3.44 MeV, respectively.
We see the fission barrier heights are significantly overestimated at large deformations with nuclear forces which are good at descriptions of nuclear masses.
It is difficult to obtain a satisfied parameterization for fission barriers based on the extended UNEDF0 force by only adjusting the momentum independent parameters.
In addition, the pairing interaction strength can also affect the fission barriers~\cite{ring2010}, which can be reduced by increasing pairing strengthes.

\begin{figure}[t]
  \begin{center}
  \includegraphics[width=0.48\textwidth]{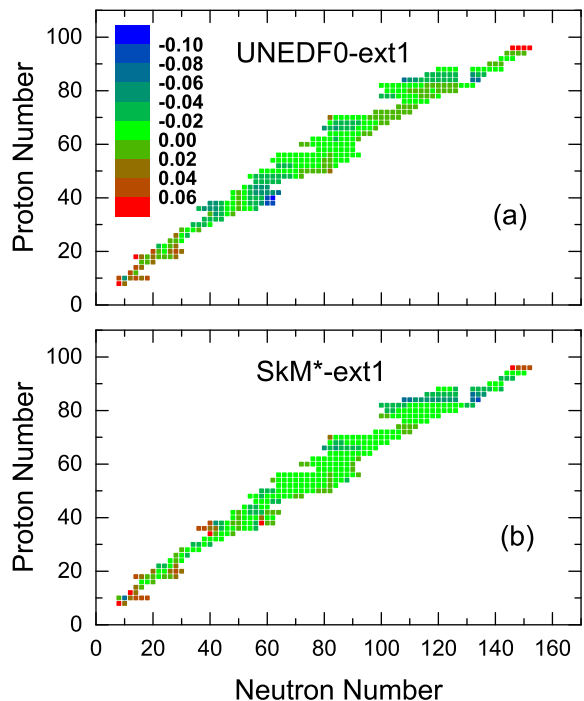}\\
  \caption{(Color online) The charge radii of 339 even-even nuclei are calculated
  by the Skyrme-Hartree-Foc+BCS method with UNEDF0$_{\rm ext1}$ and SkM$^{*}_{\rm ext1}$ forces.
  The differences between calculated values and experimental data, $R^{\rm Calc.}$-$R^{\rm Expt.}$,  are displayed.
  The unit of Charge radii is fm.
  }
  \label{fig-radii}
  \end{center}
\end{figure}

\subsection{Global charge radii}

The charge radius is also an important bulk observable associated
with nuclear saturation properties. For example, the systematic studies of charge radii
of Ca isotopes are recently a hot topic~\cite{ca}, which provides a chance to look for
the evolution of shell structures and deformations. There are extensive studies of global binding energies.
There are fewer experimental data of charge radii than that of binding energies. Fortunately the method using laser isotope shifts
is very precise for measurements of charge radii of ground states and isomeric states~\cite{ca}.

Figure \ref{fig-radii} displays the global calculations of charge radii of 339 even-even nuclei compared to experimental data~\cite{angeli}.
With the refitted UNEDF0$_{\rm ext1}$ and SkM$^{*}_{\rm ext1}$ forces, the obtained
charge radii rms are 0.027 fm and 0.023 fm respectively. Actually the original Skyrme forces and the extended Skyrme force are very close in
descriptions of charge radii.
We see that SkM$^{*}_{\rm ext1}$ descriptions of charge radii are slightly better than that of UNEDF0$_{\rm ext1}$.
Both descriptions of charge radii of light nuclei are not satisfied.
The UNEDF0 and UNEDF0$_{\rm ext1}$ forces are not good at descriptions of charge radii around $^{102}$Zr.
 Generally the descriptions of charge radii of light nuclei
are less satisfactory compared to heavy nuclei. There is not significant shell effects
in the descriptions of charge radii in contrast to binding energies.
There are several specific regions that both parametrizations can not describe well.
For example, large discrepancies are identified in $^{16}$O, $^{20,28}$Ne, $^{24}$Mg, $^{48}$Ca, $^{146,150}$Dy,  $^{192,194,216,218}$Po and $^{242-248}$Cm.
These distinct discrepancies should be considered in the future optimizations of Skyrme forces.

\subsection{Equation of state}

\begin{figure}[t]
  \begin{center}
  \includegraphics[width=0.48\textwidth]{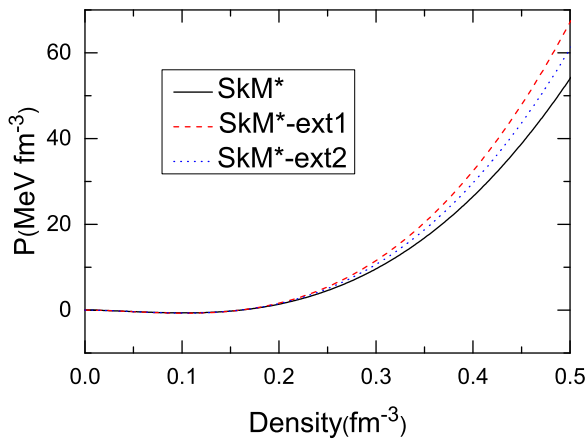}\\
  \caption{(Color online) The pressure of the symmetric  nuclear matter as a function of densities are obtained
  with SkM$^{*}$, SkM$^{*}$$_{\rm ext1}$, SkM$^{*}$$_{\rm ext2}$ forces. }
  \label{fig-p}
  \end{center}
\end{figure}
In the previous work~\cite{xiong}, we have shown that the higher-order density dependent term can
particularly affect the equation of state in the high density region.
The equation of state at the high density region is critical to address the properties of neutron stars.
Fig.\ref{fig-p} displays the pressure of symmetric nuclear matter as a function of densities, which are obtained from
the extended SkM$^{*}$ forces.
 Generally, it can be seen that the pressure from the extended forces
increased at the high density region compared to the original forces.
This is consistent with our previous results based on the extended SLy4 force~\cite{xiong}.
The differences between UNEDF0$_{\rm ext1}$, UNEDF0$_{\rm ext2}$  and UNEDF0 are very small and are not shown.
At the saturation point, the incompressibilities of the extended forces slightly increase.

\begin{figure}[t]
  \begin{center}
  \includegraphics[width=0.48\textwidth]{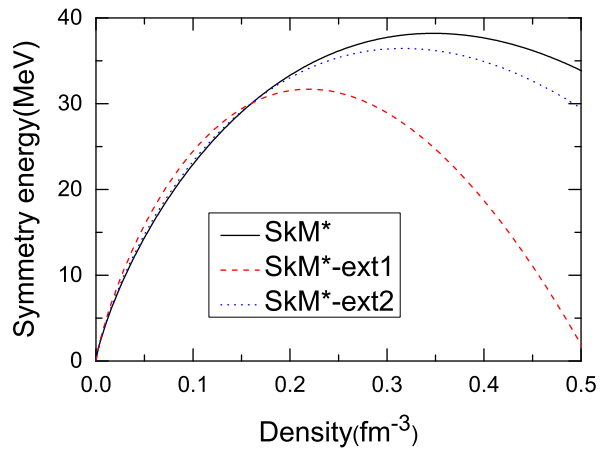}\\
  \caption{(Color online) The symmetry energy of the symmetric  nuclear matter as a function of densities are obtained
  with SkM$^{*}$, SkM$^{*}$$_{\rm ext1}$, SkM$^{*}$$_{\rm ext2}$ forces.  }
  \label{fig-sym}
  \end{center}
\end{figure}

Figure \ref{fig-sym} displays the symmetry energies as a function of densities.
Generally, it can be seen that the symmetry energies of the extended forces decrease at high density region
compared to the original forces. This is also consistent with our previous results based on SLy4~\cite{xiong}.
Note that the symmetry energy at the saturation point has not been adjusted.
Our results show that the symmetry energies at high densities consistently become soft, and this has been indicated by the experimental
$\pi^{-}/\pi^{+}$ ratio ~\cite{xiao} although soft symmetry energies at high densities are still controversial.
In both Fig.\ref{fig-p} and Fig.\ref{fig-sym}, the equation of state from SkM$^{*}$$_{\rm ext2}$  which is refitted with inputs of fission barriers are between SkM$^{*}$ and SkM$^{*}$$_{\rm ext1}$.
The symmetry energy of the extended force at the saturation point is unchanged  in the fitting procedure while the slope is changed.
In particular, the slope of symmetry energy $L$ of UNEDF0$_{\rm ext2}$($L$=51.8 MeV) is lager than that of UNEDF0 ($L$=45 MeV) and UNEDF0$_{\rm ext1}$($L$=45.4 MeV).
The slope of symmetry energy of SkM$^{*}$ ($L$=45.8 MeV)and SkM$^{*}$$_{\rm ext2}$($L$=42.5 MeV) are larger than that of SkM$^{*}$$_{\rm ext1}$($L$=27.5 MeV).
This indicates that the fitting including fission barriers tends to increase the slope of symmetry energy.

\begin{figure}[t]
  \begin{center}
  \includegraphics[width=0.48\textwidth]{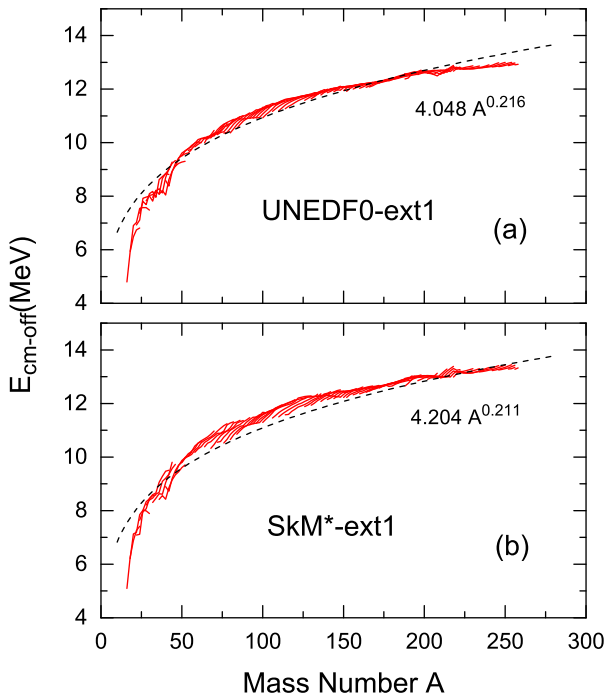}\\
  \caption{(Color online) The off-diagonal center-of-mass corrections are obtained by deformed Skyrme-Hatree-Fock+BCS calculations with  (a) UNEDF0$_{\rm ext1}$,
  (b)SkM$^{*}$$_{\rm ext1}$, respectively. The fitted functions are also given as dashed lines.   }
  \label{fig-com}
  \end{center}
\end{figure}

\subsection{Global center-of-mass corrections}

Before we develop the next-generation effective nuclear forces, we
should comprehensively understand the beyond mean-field corrections.
Symmetries in unrestricted density functional theory are spontaneously broken to account
many-body correlations. Correspondingly, there are various projection methods to restore the broken symmetries~\cite{benderrev}.
For example, the center-of-mass correction is used to restore the translation symmetry, the
angular momentum projection is used to restore the rotation symmetry,
the particle number projection is used to restore the gauge symmetry due to pairing.

The center-of-mass correction is in principle important in light nuclei and in \textit{ab initio} calculations~\cite{jvary}.
It has been demonstrated to be important for descriptions of nuclear surface properties~\cite{bender}.
On the other hand, the optimized density functional theory can give good descriptions of fission barriers~\cite{markus2}.
It is desirable to study the global center-of-mass corrections.
The center-of-mass (c.m.) correction energy includes the diagonal term (one-body) and the off-diagonal term (two-body)~\cite{bender} as:
\begin{equation}
E_{c.m.}=\frac{1}{2mA}\sum_{i=1}^{A} \mathbf{P}_i^2+\frac{1}{2mA}\sum_{i>j} \mathbf{P}_i\cdot\mathbf{P}_j
\end{equation}

Figure \ref{fig-com} displays the global off-diagonal c.m. corrections, which are actually comparable to the diagonal c.m. contributions but with a opposite sign.
Note that systematic calculations of c.m. corrections with standard Skyrme forces have been performed in Refs.~\cite{bender,reinhard}.
It can be seen that the shell effects and isospin dependencies in the c.m. corrections are not significant from both UNEDF0$_{\rm ext1}$ and SkM$^{*}$$_{\rm ext1}$ forces.
For UNEDF0$_{\rm ext1}$, the smooth two-body c.m. contributions can be refitted roughly as 4.05A$^{0.216}$.  Note that the
one-body diagonal c.m. corrections can be fitted roughly as -14.58A$^{0.047}$ which is almost mass independent. The total c.m. correction can be fitted as  -18.33A$^{-0.208}$.
For SkM$^{*}$$_{\rm ext1}$, the  one-body and two-body corrections can be fitted as -14.916A$^{0.046}$ and 4.20A$^{0.211}$ respectively, and the total c.m. correction is -18.61A$^{-0.213}$.
We see the two different Skyrme forces have very close c.m. corrections. For $A<$40, the off-diagonal c.m. corrections deviate from the fitted functions, indicating
that microscopic c.m. corrections play a special role in light nuclei.

In Ref.~\cite{bender} , it has been
pointed out that the c.m. correction is closely related to surface energies.
The role of surface energies associated with Skyrme forces has been extensively studied~\cite{32}.
 In our calculations, the two body term is a function
of  $A^{0.2}$, which is close to the surface curvature term rather than the surface term.
The inclusion of the curvature term in the liquid drop model can indeed remarkably improve
the description of fission barriers, although it is not essential for descriptions of binding energies~\cite{pomorski}.
It is still a puzzle that the curvature coefficient from the leptodermous expansion in self-consistent calculations are much larger than that in liquid drop model~\cite{reinhard,durand}.
Based on our results, \textit{a posteriori} two-body c.m. correction can significantly reduce the curvature coefficient in microscopic calculations.
We see that the one-body and two-body c.m. corrections have very different nuclear mass dependencies.
This may imply a mass-dependent nuclear force if only the one-body c.m. correction is included.
The simultaneous optimization of Skyrme forces for both binding energies and fission barriers has not been satisfied so far.
It is controversial that the fission barriers of Actinide nuclei are mainly correlated with the surface symmetry energy rather than the surface curvature energy~\cite{nikola}.
Our results demonstrated that
the off-diagonal c.m. correction is related to the surface curvature energy and thus is important for descriptions of surface properties,
which is beyond the optimizations of Skyrme forces with only diagonal c.m. corrections.

\begin{figure}[t]
  \begin{center}
  \includegraphics[width=0.48\textwidth]{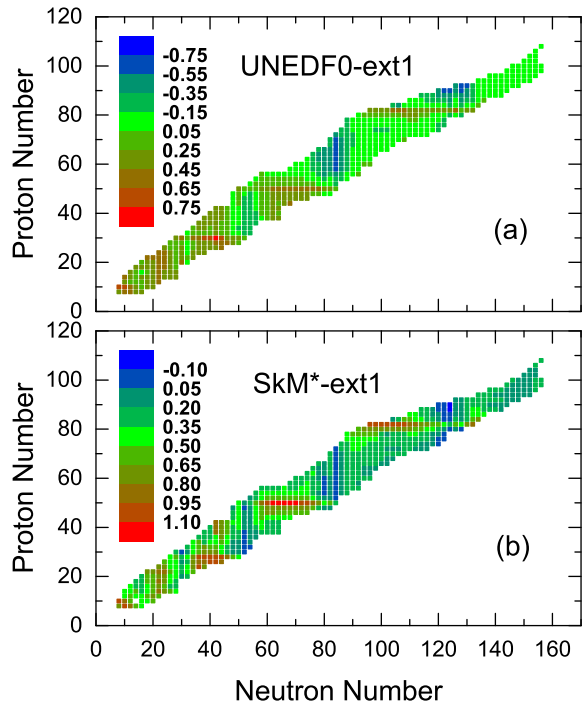}\\
  \caption{(Color online) The comparison of binding energies between Hartree-Fock-Lipkin-Nogami and Hartree-Fock-BCS calculations with different Skyrme forces,
  (a) UNEDF0$_{\rm ext1}$,
  (b)SkM$^{*}$$_{\rm ext1}$.    }
  \label{fig-ln}
  \end{center}
\end{figure}

\begin{table}
  \caption{Pairing strengthes for Skyrme Hartree-Fock calculations with UNEDF0 (and UNEDF0$_{\rm ext1}$) and SkM$^{*}$ (and SkM$^{*}_{\rm ext1}$), respectively.
  The unit of the pairing strength is MeV fm$^{3}$.  }
  \label{table2}
  \centering
  \begin{tabular}{ccccc}
  \hline
\multicolumn{1}{c}{}   & \multicolumn{2}{c}{UNEDF0}    & \multicolumn{2}{c}{SkM$^*$} \\
\cline{2-3}
\cline{4-5}

     &Proton     &Neutron        &  Proton     & Neutron       \\
\hline
HF-BCS           & 400&340 &480&450    \\
 HF-LN & 260&210    &305&275    \\
\hline
\end{tabular}
\end{table}

\subsection{Global Lipkin-Nogami corrections}

The pairing correlations in nuclei can be treated by BCS or Bogoliubov approximations in the framework of independent quasiparticle particles, which is associated with the spontaneous breaking of the gauge symmetry and the
non-conservation of particle numbers.
For simplicity, the Lipkin-Nogami method~\cite{lipkin,nogami} is usually adopted to conserve the particle numbers at the order of $(\Delta N)^2$.
To demonstrate the effects of Lipkin-Nogami (LN) corrections, we display the binding energy differences between
the Hartree-Fock-BCS (HF-BCS) and Hartree-Fock-LN (HF-LN) methods.
\begin{equation}
\Delta E_{LN} = E_{\rm HF-LN}-E_{\rm HF-BCS}
\end{equation}

Figure \ref{fig-ln} displays  the global LN corrections of $\Delta E_{LN}$, which are calculated with UNEDF0$_{\rm ext1}$ and SkM$^{*}$$_{\rm ext1}$ forces
and mixed-pairing interactions. The pairing strengthes adopted by two approaches in this work have been adjusted to reproduce the pairing gaps in
$^{252}$Fm, as listed in Table~\ref{table2}. The adjusted LN pairing strengthes are  slightly smaller than the BCS pairing strengthes.
The proton pairing strengthes are slightly larger than neutron pairing strengthes.
The pairing strengthes for UNEDF0$_{\rm ext1}$ are smaller than that in SkM$^{*}$$_{\rm ext1}$ due to a larger effective mass.
It can be seen that generally the binding energy differences are within 0.75 MeV for UNEDF0$_{\rm ext1}$ and are less than 1.1 MeV for SkM$^{*}$$_{\rm ext1}$. It is evident that the global
LN corrections are related to shell structures. The global pattern of LN corrections with the two Skyrme forces are very similar.
Compared to the BCS approximation,
the LN approximation gives more binding energies for neutron shell gaps than proton shell gaps.
For the light nuclei, the feature of LN corrections is complex.
Statistically, the earlier study has pointed out that the restoration of the exact particle number
doesn't improve significantly the global descriptions of nuclear masses~\cite{samyn}. In our calculations with UNEDF0$_{\rm ext1}$,
the rms of binding energies with the HF-LN approach is 1.291 MeV which is slightly better than the HF-BCS approach of 1.316 MeV.
For SkM$^{*}$$_{\rm ext1}$, the rms errors of HF-LN and HF-BCS are almost the same.

\section{Summary}

In summary, we have studied the global performances of Skyrme forces with an extended higher-order density dependent term.
Our studies are based on two very different Skyrme forces: UNEDF0 optimized for nuclear masses and SkM$^{*}$ optimized for fission barriers.
We only adjusted the momentum independent parameters.
The global descriptions of binding energies with UNEDF0$_{\rm ext1}$ have obtained a rms of 1.29 MeV, which is encouraging, compared
to the best-optimized UNEDF0 rms of 1.455 MeV.
In addition, the systematic analysis demonstrated that binding energies of $N=Z$ nuclei have been generally underestimated.
The descriptions of charge radii are generally good except some local regions.

We demonstrated that there is a competition in the simultaneous optimization of binding energies and fission barriers.
In this respect, our systematic calculations demonstrated that the off-diagonal center-of-mass corrections are numerically related to the surface curvature energy rather than the surface energy,
 which is important for proper description of surface properties and should be included in the future optimizations.
The features of Lipkin-Nogami pairing corrections with two Skyrme forces are very similar and are related to shell gaps.
Statistically, the Lipkin-Nogami method can not significantly improve the descriptions of global binding energies.
We have not yet studied the rotational corrections related to deformations, which involve configuration mixtures and are more complicated.
These microscopic corrections may be bridged to the phenomenological corrections in high-precision nuclear mass models~\cite{hfb14}.
We also studied the influences of the high-order density dependent term on the equation of state, which mainly impact the
 high-density properties and
are consistent with our previous study based on the SLy4 force. The higher-order density dependent term has large impacts in Skyrme forces with a small power factor $\gamma$ such as SLy4 and SkM$^{*}$ forces.
Presently we only adjusted the momentum-independent parameters based on existing Skyrme forces.
The optimization of extended Skyrme forces with all parameters are in progress and the performances are expected to be further improved.
Our global analysis should be useful for
future developments of high-precision nuclear energy density functionals.

\section{Acknowledgments}
 This work was supported by the National Natural Science Foundation of China under Grant No.11522538.
We also acknowledge that computations in this work were performed in Tianhe-1A
located in Tianjin and Tianhe-2
located in Guangzhou.


\end{document}